\documentclass[aps,prd,amssymb,showpacs,floatfix,preprint]{revtex4-1}
\usepackage{amsmath,amsfonts,graphics,epsfig,amssymb}
\begin{document}
\title{
Fermionic zero modes in the vortex field in arbitrary dimensions and index of Dirac operator with Majorana-like interaction
}
\author{G.~P.~Bednik}

\affiliation{Institute for Nuclear Research of RAS, 60th October
Anniversary Prospect 7a, Moscow 117312, Russia}
\affiliation{Physics Department, Moscow State University,
Vorobjevy Gory, 119991, Moscow, Russia}
\begin{center}

\begin{abstract}
In this work we consider fermionic zero modes in the external scalar and electromagnetic field forming the vortex on a sphere. We find the correspondence between the equations for the fermions in different dimensions, find their explicit expressions through the vortex fields in case of massless fermions, asymptotics near the poles in case of massive fermions and check the number of the solutions by proving index theorem for the fermions on a sphere. As a part of deriving the index, we write a detailed calculation of the Green function of the Heat equation.
\end{abstract}

\end{center}
\maketitle

\section{Introduction}

Recently, considerable interest has been attracted to localized fermionic
states in topologically non-trivial external fields. Its revival is
stimulated by developments in two very different areas of physics,
high-energy particle theory and condensed-matter physics (theory and
experiment).

The Standard Model of particle physics, though extremely successful in
description of most of the experimental results, cannot explain the
apparent hierarchy between two fundamental scales, the electroweak
symmetry-breaking scale $v\sim$200~GeV and the gravitational (Planck)
scale $M_{\rm Pl}\sim 10^{19}$~GeV. It has been suggested \cite{ArkaniHamed:1998vp} and
discussed in thousands of research papers (for reviews, see e.g.\
Ref.~\cite{Rubakov:2001kp, Rubakov:2003wx}) that the gravitational scale may be much
lower if there exist additional space dimensions beyond the three observed ones.
Their non-observation may be related either to compactification of extra
dimensions down to a very small size \cite{Kaluza, Klein} or to
localization of visible particles on the observed four-dimensional
Minkowski subspace \cite{Rubakov:1983bb, Visser:1985qm}. The latter may naturally happen
due to interaction of a particle with the external field with non-trivial
topological properties (e.g.\ \cite{Jackiw:1975fn, Jackiw:1981ee}).

Quite recently, it has been realized that a similar situation may occur in
a laboratory, where non-trivial topology of the effective bosonic
external field results in the localization of fermionic states at
particular points, lines or surfaces in certain materials. The motion of
fermions is governed in these cases by the very same equations as in the
particle-physics models and in many cases, the localized states have been
observed in a laboratory. Well-known examples include the fractional Hall
charges \cite{frac-Hall}, topological insulators \cite{Fu:2008zzb},
topological superconductors \cite{Nishida:2010vj} and graphene
\cite{graphene}. The topological nature of the localized states makes them
in principle suitable for the purposes of quantum computing \cite{Kitaev}.
Reviews of some topics in this rapidly developing field may be found in
Ref.~\cite{Manifestations of topological effects in graphene}.

In this paper, we will study localized fermionic states in the external
field of abelian Abrikosov--Nielsen--Olesen vortex
\cite{Abrikosov, Nielsen:1973cs}. 

The purpose of the present work is to explore various
possible fermion-vortex interactions for various number of space
dimensions. In (3+1) dimensions, it has been shown that for the vortex
winding number $n$, there exist $n$ localized zero modes of fermions with
either Majorana-like~\cite{ Jackiw:1981ee}  or Dirac-like \cite{Witten:1984eb}
equations. These zero modes describe (1+1)-dimensional massless fermions
which move freely in the core of the vortex line dubbed ``superconducting
string''; the existence of these modes is related to the vortex topology
by the index theorem \cite{Weinberg:1981eu}. This feature has been used for
extra-dimensional model building~\cite{ArkaniHamed:1998vp, Rubakov:1983bb} in (5+1) dimensions,
where the vortex fields depend on two extra spatial coordinates while the
core corresponds to our three-dimensional space. Two (related to each
other) features of the localized modes are important for construction of
realistic particle models: chirality and masslessness. The former provides
for correct quantum numbers of the standard-model fermions while the
latter opens up a possibility to generate particle masses at a correct
scale by means of an additional interaction with the Higgs scalar field
treated perturbatively. Moreover, if three fermionic generations are
associated with three zero modes of an $n=3$ vortex, then an elegant
mechanism to explain the mass hierarchy among the families of charged
fermions may operate \cite{Libanov:2000uf, Frere:2003yv}. It has been shown \cite{Libanov:2000uf,  Frere:2003yv, Libanov:2002ka}
that within this approach, not only the hierarchical pattern of masses and
mixing may be reproduced, but their particular experimental values are
obtained with a surprisingly small number of free parameters. Finally, the
very same model may automatically explain \cite{Frere:2010ah} the observed
hierarchical pattern of neutrino masses and leptonic mixings whch is very
different (e.g.\ Ref.~\cite{Hirsch:2008rp}) from the one observed among charged
particles.

Though various compactification schemes for the extra dimensions may be
considered, a practical working example \cite{Frere:2003yv} for the quoted
phenomenological results was the $M_{4}\times S_{2}$ manifold, where
$M_{4}$ is the usual (3+1)-dimensional Minkowski space and $S_{2}$ is the
extra-dimensional sphere. Though the vortex on the sphere has a somewhat different
topology from the flat space (for instance, nontrivial topological
configurations may exist for purely gauge fields), it has been
shown~\cite{Frere:2003yv} that the basic prperties of the flat-space solution
remain intact. However, to the best of our knowledge, the index theorem
for the spherical case has not been proven.

Returning to condensed-matter examples, we point out that the
vortex-on-the-sphere equations are relevant for description of fermionic
states on a fullerene molecule, where the effective vortex-like external
field is induced by disclinations in the hexagon vertices
\cite{Roy:2009vc}. Possible applications of the fullerene-localized
fermions are yet to be explored.

Motivated by these important applications, we consider here the fermions
with 
Majorana-like interaction
with the vortex field. The fermions are assumed to live on a $M \times
S_{2}$ manifold, the vortex field depending on the two $S_{2}$
coordinates. We will relate Majorana and Dirac cases for
arbitrary $D$ to each other (and ultimately to the cases of lower dimensions), study the properties of the fermionic zero modes and derive
the index theorem.

The rest of the paper is organized as follows. In Sec.\ref{ZM} we write the vortex-fermion interaction, study the fermion equations of motion
and perform the reduction of the cases of different dimensionality to each
other. We find the structure of the fermionic zero modes by a
direct analysis of these equations made in \ref{solving}. In Sec.\ref{IndexOfTheDiracOperator} we derive the corresponding
indices and prove particular index theorems which of course are in
agreement with the explicit results of Sec.\ref{ZM}. We summarize our conclusions,
emphasise properties of two phenomenologically most interesting cases
($M_{4}\times S_{2}$ -- extra-dimensional extension of the
particle-physics Standard Model ) and
discuss possible applications of other cases in the section \ref{Conclusions}. Notations and
technical details are collected in the Appendices.

\section{Structure of zero modes} \label{ZM}

In this paper we consider fermions on multi-dimensional manifold  $S^2 \times M $, where $M$ is Minkovski space. If the dimensionality of the manifold is even, we designate it by $D$. If the dimensionality is odd, it is convenient to designate it by $D+1$. The sphere contains an Abelian gauge and a scalar field which form Abrikosov-Nielsen-Olesen vortex. Its structure is described in detail in ref.\cite{Frere:2003yv}. We briefly recall it here.
The Lagrangian for these fields is:
\begin{eqnarray}
L = \sqrt{G_S} (-\frac{1}{4} F_{ab}F^{ab} +
(D_{a}\phi)^{+}D^{a}\phi - \frac{\lambda}{2} (\phi^{2} - v^{2})^2 ),     \label{VortexLagr}
\end{eqnarray}
where  $ D_{a}\phi = \partial_{a}\phi - ieA_{a}\phi $ and $F_{ab} = \partial_a A_b - \partial_b A_a$ (all kinds of subscripts are described in appendix \ref{Notations}).
To describe the solution we have to introduce two patches (this discourse was proposed in \cite{Wu:1975es} ). The first of them covers the sphere with the South Pole excluded. In this patch one may use the following Ansatz for the solution:
\begin{eqnarray}
  \phi &=& f(\theta) e^{i\varphi}  , \nonumber\\
 A_{\varphi} &=& A(\theta)  ,\nonumber\\
  A_{\theta} &=& 0  .
\label{Eqn/Pg1/1:kurs}
\end{eqnarray}
The other patch covers the sphere with the North Pole excluded. In the overlapping region its field values are connected to the values on the first patch by the gauge transformation with the gauge function equal to $ (
-\varphi) $ , 
\begin{eqnarray}
 A \longrightarrow A - \frac{1}{e},  \qquad
 \phi \longrightarrow \phi e^{-i\varphi}.
\end{eqnarray}
It can be shown that $f(\theta)$ and $A(\theta)$ have the following asymptotics:

  \begin{eqnarray}
  \theta\rightarrow 0 :    \qquad      f &\sim& C_{0F}\theta  , \quad
A \sim C_{0A}\theta^{2}, \nonumber\\
\theta\rightarrow\pi :  \qquad   f &\sim& C_{\pi F}  , \quad   A\sim\frac{1}{e}  ,
\label{VortexAsymp}
\end{eqnarray} 
where $ C_{0F}$, $C_{0A}$, $ C_{\pi F}$ are constants.

Now let us consider fermions interacting with the vortex. To construct the Lagrangian we introduce velbain $e^{A}_{\alpha}$ which is: 
\begin{eqnarray}
e_{\theta}^{\alpha} &=& \frac{1}{R} \delta_{\theta}^{\alpha}, \nonumber\\
e_{\varphi}^{\alpha} &=& \frac{1}{R\sin\theta} \delta_{\varphi}^{\alpha}, \nonumber\\    
e_{\mu}^{\alpha} &=& \delta_{\mu}^{\alpha}  \label{velbain}
\end{eqnarray}
Then we define the spin connection $ R^{\alpha \beta}_{B}= \frac{1}{2} e_{A}^{\alpha}\nabla_{B}e^{A,\beta}$.
It can be shown that $R^{12}_2 = - R^{21}_2 =  \frac{\cos \theta}{2} $ and all other components are equal to zero.
 The Lagrangian for the fermions is defined as
\begin{eqnarray}
L = \sqrt{-G}  \left( i \bar{N} \Gamma^{\alpha}_{(D)} e^{A}_{\alpha} D_{A} N  + (\frac{g}{2} \Phi^{2k}N^+ \Gamma^0_{(D)} C_{(D)} N^* + h.c. )\right), 
\end{eqnarray}

where
\begin{eqnarray}
 D_{A} = \partial_{A} + \frac{ R^{\beta \gamma}_{A}}{2} \Gamma^{\beta}_{(D)} \Gamma^{\gamma}_{(D)} - ie(k+\kappa \tilde \Gamma^{D}_{(D)})A_{A},
\end{eqnarray}
$g$ can be chosen a real coupling constant without loss of generality (if $g$ had a complex phase, we could include it into $N$ ), $k$ and $\kappa$ are integer or half-integer charges.
So in case we consider the Lagrangian is:

\begin{eqnarray}
L = \sqrt{-G}  \left(  i \bar{N} \Gamma^{\mu}_{(D)}  \partial_{\mu} N + \right. 
i \bar{N}\frac{\Gamma^1_{(D)}}{R} (\partial_{\theta}+ \frac{\cot \theta}{2})N + \nonumber\\ 
 i\bar N \frac{\Gamma^2_{(D)}}{R\sin \theta} (\partial_{\varphi} - ieA_{\varphi} (k+ \kappa \tilde \Gamma^{D}_{(D)}) )N + 
  \frac{g}{2} \left. (\Phi^{2k}N^+ \Gamma^0_{(D)} C_{(D)} N^* + h.c. )\right) \label{L_N}
\end{eqnarray}

Let us note that since $N$ is a column of anticommuting variables, the term responsible for the  interaction with the scalar field in (\ref{L_N}) is non-zero only if  $\Gamma^0_{(D)} C_{(D)}$ is antisymmetric. If we take into account (\ref{(Gamma^0_{(D)} C_{(D)})^T}) we see that  in even number of dimensions a few cases are possible:
\begin{itemize}
\item{D/2 mod 4 = 0,} no term with majorana mass
\item{D/2 mod 4 = 1,}  (a) $C_{(D)}=C^{(1)}_{(D)}$
\item{D/2 mod 4 = 2,}  (b) $C=C^{(2)}_{(D)}$, (c) $C=C^{(1)}_{(D)}$
\item{D/2 mod 4 = 3,}  (d) $C=C^{(2)}_{(D)}$
\end{itemize}
Also if we consider a gauge transformation of the term $ \frac{g}{2} \Phi^{2k}N^+ \Gamma^0_{(D)} C_{(D)} N^*$, we can see that the term containing $\kappa$ is allowed by the gauge symmetry only if the anticommutator $\{ \Gamma^{D}_{(D)} , \Gamma^0_{(D)} C_{(D)} \} = 0$, that is in cases (a),(d).

If the number of dimensions is odd and equal to $D+1$, to obtain Lorentz-invariant couplings we cannot assume $C= C^{(2)}_{(D)}$ or write the term containing $\kappa$.  
So we are left with the following cases:
\begin{itemize}
\item{D/2 mod 4 = 0 or 3,} No term with Majorana mass
\item{D/2 mod 4 = 1,} $C_{(D)}=C^{(1)}_{(D)}$
\item{D/2 mod 4 = 2,} $C_{(D)}=C^{(1)}_{(D)}$
\end{itemize}
Thereby the equations for odd number of dimensions can be considered as a special case of the equations for even number of dimensions except for $D=2$.

The Lagrangian results in the following equations of motion:
\begin{eqnarray}
i\Gamma^{\mu}_{(D)}\partial_{\mu}N + 
   i\frac{ \Gamma^1_{(D)}}{R} (\partial_{\theta}+ \frac{\cot \theta}{2})N +\nonumber\\
 i\frac{\Gamma^2_{(D)}}{R\sin \theta} (\partial_{\varphi} - ieA_{\varphi} (k+ \kappa \tilde \Gamma^{D}_{(D)}) )N +
g \Phi^{2k} C_{(D)} N^* =0 \label{rawEOM}
\end{eqnarray}
$N$ can always be decomposed into a set of plane waves propagating in $M$. Separating positive and negative frequency, we get the following expression:
\begin{eqnarray} 
 N(x^{A}) = \int d^4 k (   N_+ (k_{\mu} , x_a)  e^{ik_{\mu}x^{\mu} }   +       
 N_- (k_{\mu} , x_a)  e^{-ik_{\mu}x^{\mu} }  ) .
\end{eqnarray}
After substituting this decomposition into eq.(\ref{rawEOM}),  multiplying it by $\Gamma^0$ and separating the terms with different exponents we have the following equations:
\begin{eqnarray}
(-k_0 - \Gamma^0_{(D)} \Gamma^i_{(D)} k_i) N_+ + D_{kin}N_+ +g \Phi^{2k} \Gamma^0_{(D)} C_{(D)} N_-^* = 0, \nonumber\\
( k_0 + \Gamma^0_{(D)} \Gamma^i_{(D)} k_i) N_- + D_{kin}N_-   +g \Phi^{2k} \Gamma^0_{(D)} C_{(D)} N_+^* = 0, \label{N+-}
\end{eqnarray}
where
\begin{eqnarray}
 D_{kin}   \equiv 
  i\frac{\Gamma^0_{(D)} \Gamma^1_{(D)}}{R} (\partial_{\theta}+ \frac{\cot \theta}{2}) + i\frac{\Gamma^0_{(D)}\Gamma^2_{(D)}}{R\sin \theta} (\partial_{\varphi} - ieA_{\varphi} (k+ \kappa \tilde \Gamma^{D}_{(D)}) ).
\end{eqnarray}
We introduce a bispinor
\begin{eqnarray}
 \tilde{N} =
 \left(\begin{array}{c}
N_+  \\  N_-^{*}
\end{array}\right) \label{NN^*}
\end{eqnarray} 
and rewrite the equations of motion as
\begin{eqnarray}  
 k^0 
\tilde{N}    =
 \tilde{A^i} k^i\tilde{N} +
D\tilde{N}  , \label{EqtildeN}
\end{eqnarray}
where
\begin{eqnarray}
 \tilde{A^i} &=& 
 \left(\begin{array}{cc}
\Gamma^0_{(D)} \Gamma^i_{(D)}  & 0 \\
0  &  (\Gamma^0_{(D)} \Gamma^i_{(D)} )^*
\end{array}\right)   \nonumber
\end{eqnarray}
and
\begin{eqnarray}
 D &=& 
 \left(\begin{array}{cc}
D_{kin}       &     g\Phi^{2k} \Gamma^0_{(D)} C_{(D)}   \\
- g(\Phi^*)^{2k} (\Gamma^0_{(D)} C_{(D)})^*      &    - D_{kin}^*  
\end{array}\right). \label{DiracOper}
\end{eqnarray}

One can note the following anticommutation property of $\tilde{A^i}$:
\begin{eqnarray}
\{ \tilde{A^i}  \tilde{A^j}  \}  &=&  2 \delta^{ij},  \nonumber\\
\{  A^i   D  \}   &=&   0.
\end{eqnarray}
By making use of the last relation, we conclude that if $\tilde{N}$ is an eigenvector of $D$ with an eigenvalue $m$:
$$  D \tilde{N} = m \tilde{N},  $$
then $\tilde{N_A} = \frac{k^i A^i}{|k|} \tilde{N}  $, where $|k| = k^i k^i$ , is also an eigenvector of $D$ but with opposite eigenvalue,
$$  D \tilde{N_A} = - m \tilde{N_A}.  $$
Also one can see that if $\tilde{N}$ satisfies Eq. (\ref{EqtildeN}), $\tilde{N_A}$ also satisfy this equation.
These relations can be written in the following form:
\begin{eqnarray}
k^0
 \left(\begin{array}{cc}
\tilde{N}  \\ \tilde{N_A}
\end{array}\right) 
=
 \left(\begin{array}{cc}
m & |k|  \\
|k| & -m
\end{array}\right) 
 \left(\begin{array}{cc}
\tilde{N}  \\ \tilde{N_A}
\end{array}\right). 
\end{eqnarray} 

This system has non-trivial solutions when $(k^0)^2 =  (k^i)^2 + m^2$. In this paper we will explore the zero modes, that are  eigenvectors with $m=0$.

So let us consider the equation $D \tilde{N} = 0$ which can also be written as
\begin{eqnarray}
D_{kin}  \frac{N_+ + N_-}{2} + g\Phi^{2k} \Gamma^0_{(D)} C_{(D)}  \left( \frac{N_+ + N_-}{2}\right)^* = 0, \nonumber\\
D_{kin} \frac{N_+ - N_-}{2i} + g\Phi^{2k} \Gamma^0_{(D)} C_{(D)}  \left(\frac{N_+ - N_-}{2i}\right)^* = 0. 
\end{eqnarray}
We see that these two equations have similar structure. Further we will show that they have a finite number of fundamental solutions $N^{(l)}$. If we find them,  we can write a general solution as: 
\begin{eqnarray} 
\frac{N_+ + N_-}{2} =  \sum_{l} \alpha_l N^{(l)},  \nonumber\\
\frac{N_+ - N_-}{2i} =  \sum_{l} \beta_l N^{(l)}, 
\end{eqnarray}
or
\begin{eqnarray}
N_+ = (\alpha_ l + i \beta_l) N^{(l)},  \\
N_- = (\alpha_ l - i \beta_l) N^{(l)}.
\end{eqnarray}
Here $\alpha_l$ and $\beta_l$ are arbitrary anticommutative coefficients. They are real because the equations are linear only in terms of multiplying by real numbers. Also let us note that since $\alpha_l$,$\beta_l$ are anticommutative $N^{(l)}$ are commutative. From the last formulae we conclude that $N_+$ have the same structure as $N_-^*$.

Now let us consider the fundamental solutions. We will reduce the higher-dimensional equations to the case of two dimensions. The asymptotics of the solutions are discussed in the appendix \ref{solving}. The equation for the fundamental solutions is: 
\begin{eqnarray}
i \Gamma^1_{(D)} (\partial_{\theta}+ \frac{\cot \theta}{2})N^{(l)} + i\frac{\Gamma^2_{(D)}}{\sin \theta} (\partial_{\varphi} - ieA_{\varphi} (k+ \kappa \tilde \Gamma^{D}_{(D)}) )N^{(l)} +\nonumber\\
gR \Phi^{2k} C_{(D)} (N^{(l)})^* =0 \label{EOM}
\end{eqnarray}
We note that inside the vortex $N^{(l)}$ obeys the effective Dirac equation in $M$,
\begin{eqnarray}
 i\Gamma^{\mu}_{(D)}\partial_{\mu} N= 0. \nonumber
\end{eqnarray}

A $2^{D/2}$-component spinor $N^{(l)}$ can be represented as a column consisting of two $(2^{D/2-1})$-component spinors $N_1, N_2$:
\begin{eqnarray}
N^{(l)}=
\left(\begin{array}{c}
 N_1 \\ N_2  \label{N_12}
  \end{array}\right). 
\end{eqnarray}
 Let us substitute this expression for $N^{(l)}$ into the eq. (\ref{EOM}). The result is different for each case so  let us consider them separately.

In case (b) Eqs.(\ref{EOM}) decouple into two equivalent systems:
\begin{eqnarray}
i \Gamma^1_{(D-2)} (\partial_{\theta}+ \frac{\cot \theta}{2})N_{1,2} + i\frac{\Gamma^2_{(D-2)}}{\sin \theta} (\partial_{\varphi} - ieA_{\varphi} k )N_{1,2} +
gR \Phi^{2k} C^{(1)}_{(D-2)} N_{1,2}^* =0.  \label{EOMb}
\end{eqnarray}

In case (c), the equations decouple into the systems 
\begin{eqnarray}
i \Gamma^1_{(D-2)} (\partial_{\theta}+ \frac{\cot \theta}{2})N_{1} + i\frac{\Gamma^2_{(D-2)}}{\sin \theta} (\partial_{\varphi} - ieA_{\varphi} k )N_{1} +
igR \Phi^{2k} C^{(1)}_{(D-2)} N_{1}^* =0, \nonumber\\
i \Gamma^1_{(D-2)} (\partial_{\theta}+ \frac{\cot \theta}{2})N_{2} + i\frac{\Gamma^2_{(D-2)}}{\sin \theta} (\partial_{\varphi} - ieA_{\varphi} k )N_{2} -
igR \Phi^{2k} C^{(1)}_{(D-2)} N_{2}^* =0.
\end{eqnarray}
By the phase transformation
\begin{eqnarray}
N_1 \to e^{ i \pi /4} N_1, \qquad
N_2 \to e^{-i \pi /4} N_2
\end{eqnarray}
we reduce the equations to the same form as in case (b).

We can see that Eqs. (\ref{EOMb}) have the same form as the equations of motion (\ref{EOM}) for the fermions in $(D-1)$ or $(D-2)$-dimensional space. 
So both cases (b) and (c) are reduced to the case (a). 
 
Now let us consider $D$-dimensional fermions in case (a). 
After the substitution (\ref{N_12}) the equations become: 
 \begin{eqnarray}
i \Gamma^1_{(D-2)} (\partial_{\theta}+ \frac{\cot \theta}{2})N_{1} + i\frac{\Gamma^2_{(D-2)}}{\sin \theta} (\partial_{\varphi} - ieA_{\varphi} (k+\kappa) )N_{1} -
igR \Phi^{2k} C^{(1)}_{(D-2)} N_{2}^* =0,  \nonumber\\
i \Gamma^1_{(D-2)} (\partial_{\theta}+ \frac{\cot \theta}{2})N_{2} + i\frac{\Gamma^2_{(D-2)}}{\sin \theta} (\partial_{\varphi} - ieA_{\varphi} (k-\kappa) )N_{2} +
igR \Phi^{2k} C^{(1)}_{(D-2)} N_{1}^* =0.
\end{eqnarray}
Let us multiply the second equation by $C^{(1)}_{(D-2)}$ and complex conjugate it.  
Then we introduce a new variable 
\begin{eqnarray}
 \tilde N_2 = i C^{(1)}_{(D-2)} N_2^* \label{N_2ToTildeN_2}
\end{eqnarray}
and obtain the following equations: 
 \begin{eqnarray}
 i \Gamma^1_{(D-2)} (\partial_{\theta}+ \frac{\cot \theta}{2})N_{1} + i\frac{\Gamma^2_{(D-2)}}{\sin \theta} (\partial_{\varphi} - ieA_{\varphi} (k+\kappa) )N_{1}  - gR \Phi^{2k} \tilde N_{2} =0,  \nonumber\\
 i \Gamma^1_{(D-2)} (\partial_{\theta}+ \frac{\cot \theta}{2})\tilde N_{2} + i\frac{\Gamma^2_{(D-2)}}{\sin \theta} (\partial_{\varphi} + ieA_{\varphi} (k-\kappa) )\tilde N_{2} - gR (\Phi^*)^{2k}  N_{1} =0.
\end{eqnarray} 
We now introduce a new spinor $\Psi$ defined as 
\begin{eqnarray}
\Psi =  \left(\begin{array}{c}
N_1\\
\tilde N_2 
\end{array}\right)   \label{N1tildeN2}
\end{eqnarray}
It describes zero modes of Dirac-like interacting fermions with the Lagrangian
\begin{eqnarray}
L = \sqrt{-G} \left(
\bar{\Psi}\Gamma^{\mu}_{(D)}\partial _{\mu}\Psi + 
\bar{\Psi} i \frac{ \Gamma^1_{(D)}}{R} (\partial_{\theta} + \frac{\cot \theta}{2} ) \Psi + 
\right.\nonumber\\  
i\bar{\Psi} \frac{\Gamma^2_{(D)}}{R \sin \theta} (\partial_{\varphi} -ieA_{\varphi} (\kappa+ k \tilde \Gamma^{D}_{(D)} ) ) \Psi - 
\nonumber\\   \left.
g \Phi^{2k}\bar{\Psi} \frac{(1- \tilde \Gamma^{D}_{(D)})}{2} \Psi - (\Phi^*)^{2k} \bar{\Psi} \frac{(1+ \tilde \Gamma^{D}_{(D)})}{2} \Psi \right)   
\end{eqnarray} 
and the equations for zero modes
\begin{eqnarray}
 i  \Gamma^1_{(D)} (\partial_{\theta} + \frac{\cot \theta}{2} ) \Psi +
i \frac{\Gamma^2_{(D)}}{ \sin \theta} (\partial_{\varphi} -ieA_{\varphi} (\kappa+ k \tilde \Gamma^{D}_{(D)} ) ) \Psi -  \nonumber\\
gR \Phi^{2k} \frac{(1-  \tilde \Gamma^{D}_{(D)})}{2} \Psi +
gR (\Phi^*)^{2k} \frac{(1+  \tilde \Gamma^{D}_{(D)})}{2} \Psi  = 0.   
\end{eqnarray} 

In case (d) we again make the change (\ref{N_2ToTildeN_2}).
Then we repeat the steps made above, make the same change (\ref{N1tildeN2}) and obtain the equations:
\begin{eqnarray}
 i  \Gamma^1_{(D)} (\partial_{\theta} + \frac{\cot \theta}{2} ) \Psi +
i \frac{\Gamma^2_{(D)}}{ \sin \theta} (\partial_{\varphi} -ieA_{\varphi} (\kappa+ k \tilde \Gamma^{D}_{(D)} ) ) \Psi -  \nonumber\\
gR \Phi^{2k} \frac{(1-  \tilde \Gamma^{D}_{(D)})}{2} \Psi -
gR (\Phi^*)^{2k} \frac{(1+  \tilde \Gamma^{D}_{(D)})}{2} \Psi  = 0.     \label{DiracEq}
\end{eqnarray} 

 Now let us reduce the system (\ref{DiracEq}) to the case $D=4$. If we represent $\Psi$ as a set of  $N$ two component spinors $\Psi_1 \ldots \Psi_N $ where $N$ is equal to $ 2^{\frac{D-2}{2}} $:
 \begin{eqnarray}
\Psi = 
\left(\begin{array}{ccc}
\Psi_1 & \ldots &  \Psi_N  \nonumber
  \end{array}\right)^T 
\end{eqnarray}
and use (\ref{Gamma_matrices_induction}),
 then the system  (\ref{DiracEq}) decouples into the following set of equations:
\begin{eqnarray}
 i \Gamma^1_{(2)} (\partial_{\theta}+ \frac{\cot \theta}{2})\Psi_{n} + i\frac{\Gamma^2_{(2)}}{\sin \theta} (\partial_{\varphi} - ieA_{\varphi} (k+\kappa) )\Psi_{n}  - gR \Phi^{2k} \Psi_{N-n+1} =0, \nonumber\\
 i \Gamma^1_{(2)} (\partial_{\theta}+ \frac{\cot \theta}{2})\Psi_{N-n+1} + i\frac{\Gamma^2_{(2)}}{\sin \theta} (\partial_{\varphi} - ieA_{\varphi} (\kappa -k) )\Psi_{N-n+1} -
gR (\Phi^*)^{2k} \Psi_{n} =0,\label{6Dsystem}\\
n = 1 ... N/2. \nonumber
\end{eqnarray}
One can see that the spinor
 \begin{eqnarray}
\left(\begin{array}{ccc}
\Psi_n \\
 \Psi_{N-n+1}  
  \end{array}\right) \nonumber
\end{eqnarray}
 satisfies  Eq.(\ref{DiracEq}) for $D=4$.


In the same way we can make a reduction from four to two dimensions. We write $\Psi_n$ and $\Psi_{N+1-n}$ in the component form:
\begin{eqnarray}
\Psi_{n} = 
\left(\begin{array}{c}
\xi_n^1    \\
\xi_n^2
\end{array}\right), 
\qquad
\Psi_{N+1-n} = 
\left(\begin{array}{c}
\eta^1_{N+1-n}    \\
\eta^2_{N+1-n}
\end{array}\right)
\end{eqnarray}
and after using (\ref{2D_gamma_matrices}) rewrite Eqs. (\ref{6Dsystem}) as 

\begin{eqnarray}
i \Gamma^1_{(2)} (\partial_{\theta} + \frac{\cot \theta}{2})
\left(\begin{array}{c}
\xi_n^1    \\
\eta^2_{N+1-n}
\end{array}\right) 
+
\frac{i}{\sin \theta} \Gamma^2_{(2)} (\partial_{\varphi} - ie (\kappa + k  \Gamma^0_{(2)})A_{2})
\left(\begin{array}{c}
\xi_n^1    \\
\eta^2_{N+1-n}
\end{array}\right) 
-\nonumber\\
gR 
\left(  \frac{1- \Gamma^0_{(2)}}{2} \Phi^{2k}   +    \frac{1+ \Gamma^0_{(2)}}{2} (\Phi^*)^{2k}  \right)
\left(\begin{array}{c}
\xi_n^1    \\
\eta^2_{N+1-n}
\end{array}\right) 
=0,\label{2D_Dirac_eq_1}
\end{eqnarray}

\begin{eqnarray}
i \Gamma^1_{(2)}  (\partial_{\theta} + \frac{\cot \theta}{2})
\left(\begin{array}{c}
\eta^1_{N+1-n}    \\
\xi^2_{n}
\end{array}\right) 
+
\frac{i}{\sin \theta} \Gamma^2_{(2)} (\partial_{\varphi} - ie (\kappa - k  \Gamma^0_{(2)})A_{2})
\left(\begin{array}{c}
\eta^1_{N+1-n}    \\
\xi^2_{n}
\end{array}\right) 
-\nonumber\\
gR 
\left(  \frac{1+ \Gamma^0_{(2)}}{2} \Phi^{2k}   +    \frac{1- \Gamma^0_{(2)}}{2} (\Phi^*)^{2k}  \right) 
\left(\begin{array}{c}
\eta^1_{N+1-n}    \\
\xi^2_{n}
\end{array}\right) 
=0.\label{2D_Dirac_eq_2}
\end{eqnarray}
One can see that the first of these equations is equivalent to (\ref{DiracEq}) in two-dimensional case, and the second one becomes equivalent after the change $k \to -k , \Phi \to \Phi^*$. In the appendix \ref{solving} the component form of these equations is given.

Thus we showed that the $D$ dimensional spinor satisfying the Dirac equation for zero modes (\ref{DiracEq}) can be expressed in terms of a 2-component spinor satisfying the Dirac equation. The structure of fundamental solution is the following:
\begin{eqnarray}
\Psi =  \left(\begin{array}{cc}
\psi_1 \\ \psi_2
\end{array}\right),  \label{psi1psi2}
\end{eqnarray}
where
\begin{eqnarray}
\psi_1 = \left(\begin{array}{cc}
0 \\ \vdots \\  \Psi_n \\ \vdots \\  0
\end{array}\right),
 \qquad
\psi_2 =  \left(\begin{array}{cc}
0  \\ \vdots \\    \Psi_{N+1-n} \\ \vdots \\ 0
\end{array}\right), \nonumber
\\
{\Psi_n =
 \left(\begin{array}{cc}
\xi_n^1 \\ \xi_n^2 
\end{array}\right),  
\qquad
\Psi_{N+1-n} =
 \left(\begin{array}{cc}
 \eta_{N+1-n}^1 \\ \eta_{N+1-n}^2
\end{array}\right)},
 \quad n = 1,...,N/2.  \label{FinalSolutions}
\end{eqnarray}
 For $g \ne 0$ the only non-zero components are  $\xi^1_n ,  \eta^2_{N+1-n} $. If $g = 0$,  non-zero components are linearly independent but their structure is more complicated and it is given in the table \ref{table1}. Explicit expressions and asymptotics of the solutions are found in the appendix ~\ref{solving}. The spinor $N^{(l)}$ has the following structure:
\begin{itemize}
\item{case (a),} 
\begin{eqnarray}
N^{(l)} =  \left(\begin{array}{c}
\psi_1\\
-i C^{(2)}_{(D-2)} \psi_2^* 
\end{array}\right);   
\end{eqnarray}
\item{case (b),}
\begin{eqnarray}
N^{(l)} =  \left(\begin{array}{c}
N^{(l)}_{(a)}  \\  0
\end{array}\right),   
\qquad
N^{(l)} =  \left(\begin{array}{c}
0 \\ N^{(l)}_{(a)}
\end{array}\right);   
\end{eqnarray}
\item{case (c),}
\begin{eqnarray}
N^{(l)} =  \left(\begin{array}{c}
e^{-\frac{i\pi}{4}} N^{(l)}_{(a)}  \\  0
\end{array}\right),   
\qquad
N^{(l)} =  \left(\begin{array}{c}
0 \\ e^{\frac{i\pi}{4}} N^{(l)}_{(a)}
\end{array}\right);   
\end{eqnarray}
\item{case (d),}
\begin{eqnarray}
N^{(l)} =  \left(\begin{array}{c}
\psi_1\\
 C^{(2)}_{(D-2)} \psi_2^* 
\end{array}\right).   \label{NlAns}
\end{eqnarray}
\end{itemize}

\begin{table}
\begin{tabular}{ | r | r | r | r | r | r | r | r | }
\hline             &  \multicolumn{5}{|c|}{\textbf{ number of solutions}} &  \multicolumn{2}{|c|}{\textbf{eigenvalue of}} \\ 
 \cline{2-8} component    & $\kappa < -k$ & $\kappa = -k$ & $-k  < \kappa < k$ & $\kappa = k$ & $\kappa > k $  &    $K^0$  &  $\tilde{K^0}$ \footnote{$K^0$ , $\tilde{K^0}$ are defined in the next section} \\
\hline
$\xi^1_1$    &                     &     & $k+\kappa$ & $2k$ & $k+\kappa$  & 1  & 1    \\
$\xi^2_1$    & $-\kappa-k$ &     &                     &          &                      & -1 &-1    \\
$\eta^2_2$ & $k-\kappa$  & $2k$ & $k-\kappa$  &       &                      & 1 & -1  \\
$\eta^1_2$ &                       &     &                       &       & $\kappa-k$   & -1 & 1\\
\hline 
\end{tabular}
\caption{The number of non-zero components in case of different $\kappa$} 
 \label{table1}
\end{table}
\footnotetext{ $K^0$,  $\tilde{K^0}$ are defined in (\ref{chirality}).}


Finally let us note that though fermions on two-dimensional sphere without time coordinate obey to the Dirac equation (\ref{EOM}), they are defined by Lagrangian different from (\ref{L_N}) because in the case of two dimensions there are no boost transformations:
 \begin{eqnarray}
L = \sqrt{-G_S}  \left( i N^+ \Gamma^{\alpha}_{(2)} e^{a}_{\alpha} D_{a} N  + \frac{g}{2} (\Phi^{2k}N^+ C_{(2)} N^* + h.c. )\right), 
\end{eqnarray}
where
\begin{eqnarray}
 D_{A} = \partial_{A} + \frac{ R^{\beta \gamma}_{A}}{2} \Gamma^{\beta}_{(2)} \Gamma^{\gamma}_{(2)} - ie(k+\kappa \Gamma^3_{(2)}) A_{A},\nonumber
\end{eqnarray}
and $e^a_{\alpha}$ is defined in the same way as (\ref{velbain}). One can show that this Lagrangian produce the same equations of motion as (\ref{EOM}).


\section{Index of the Dirac operator} \label{IndexOfTheDiracOperator}

In this section we will calculate the index of the operator$D$ defined in section ~\ref{ZM}~  by Eq.(\ref{DiracOper}). Indices of Dirac operators without Majorana-like interaction have been considered previously (see e.g. \cite{Vassilevich:2003xt}, \cite{Schwartz}). The case of the Majorana-like interaction is more involved because the Dirac operator mixes $N$ and $N^*$. For the two-dimensional flat space this problem has been solved in \cite{Weinberg:1981eu}. We consider the case of a sphere.

 Let us introduce operators of chirality which distinguish left-handed and right-handed modes.
In the case $g=0$ (no Majorana-like interaction) they are:
\begin{eqnarray}
K^0 = i\Gamma^1_{(D)} \Gamma^2_{(D)} = {\rm diag}
 \left(\begin{array}{ccc}\sigma^3 & \ldots & \sigma^3 \end{array}\right), \\
\tilde{K^0}=
\tilde \Gamma^{D}_{(D)} K^0. 
 \label{chirality}
\end{eqnarray}

These operators anticommute with the operator $ D_{kin}$. In case $g \ne 0$ we consider spinors ~(\ref{NN^*}) having doubled number of components and want to generalize these operators to save their anticommutation with $ D$. To achieve this we introduce an analog of $K^0$  defined as
\begin{eqnarray}
K = 
\left(\begin{array}{cc}
K^0 & 0 \\
0 & K^0 
\end{array}\right).
\end{eqnarray}
The analog of $\tilde{K^0}$ is
\begin{eqnarray}
 \tilde{K} = 
\left(\begin{array}{cc}
\tilde K^0 & 0 \\
0 & \tilde K^0 
\end{array}\right)
 \quad \mbox{in cases (b), (c)}\\
 \tilde{K} = 
\left(\begin{array}{cc}
\tilde K^0 & 0 \\
0 & -\tilde K^0 
\end{array}\right) \quad \mbox{in cases (a),(d)}
 \label{tildeK}
\end{eqnarray}
Now let us consider an arbitrary Dirak Hamiltonian $ D$ (it does not have to have a form considered in this paper) and a chirality operator $K$ anticommmuting with $D$. The fact that $ \{ D , K \} = 0$  implies that   $D$ transforms left-handed spinors (eigenvectors of $K$ with eigenvalue equal to 1 ) into right-handed ones  (eigenvectors of $K$ with eigenvalue equal to -1 )  and vice versa because
\begin{eqnarray}
(1+K) D =  D (1-K),\\
(1-K)  D = D (1+K).
\end{eqnarray}
Let us introduce 
\begin{eqnarray}
D_U = D \frac{(1+K)}{2}, \\
D_L =   D \frac{(1-K)}{2}.
\end{eqnarray}
It is easy to check that if $ D$ and $K$ are Hermitian, $D_U$ and $D_L$ are mutually conjugated:
\begin{eqnarray}
D_U^+ = D_L, \\
D_L^+ = D_U.
\end{eqnarray}
Therefore, if we have a Dirac Hamiltonian in the space of left-handed fermions, its Hermit conjugated operator is the Dirac Hamiltonian in the space of right-handed fermions. 

Now let us define the index as
 $${\tt index} = {\tt dim(ker}D_{L} D_{U}) -  {\tt dim(ker}D_{U} D_{L})  = {\tt tr} {\rm e}^{-t(D_{U} D_{L})} -{\tt tr} {\rm e}^{-t(D_{L} D_{U})}.  $$
This expression can be transformed in the following way:
\begin{eqnarray}
{\tt index} = 
{\tt tr} \left( {\rm e}^{-t D^2 \frac{1-K}{2}}  -  {\rm e}^{-t D^2 \frac{1+K}{2}}  \right) = \nonumber\\
{\tt tr} \left( {\rm e} ^{-t D^2} \frac{1-K}{2} -  {\rm e} ^{-t D^2}  \frac{1+K}{2}\right) =
-{\tt tr} ({\rm e} ^{-t D^2} K).
\end{eqnarray} 
Here we made use of a general property of any chirality operator $K^2 =1$.

 To find ${\rm e} ^{-t D^2} $ we consider an equation 
\begin{eqnarray}
 \frac{d}{dt} G(x,y,t) = - D^2  G ,  \qquad  G  \rightarrow \delta(x-y) \quad \mbox{at} \quad  t\rightarrow 0 
 \label{HeatEq}
\end{eqnarray}
which is known as the Heat equation; $ G $ is its Green function. According to our notations, in case $g=0$  $ G $ is a $2^{D/2}$*$2^{D/2}$ matrix and in case $g \ne 0$ it is $2^{D/2+1}$*$2^{D/2+1}$ matrix.

To continue the calculation of $G$, we introduce a scalar product in the space where $D$ acts. In sec.  ~\ref{ZM}~ we defined this space as a set of spinors having a form (\ref{NN^*}). Now we consider $D$ in the linear span of fundamental solutions of  (\ref{NN^*}) with commutative coefficients. We have to do this because if the coefficients were anticommutative, the scalar square of any element would be equal to zero. Certainly the change of coefficients from anticommutative to commutative does not affect the equations since they are linear, and hence does not affect the index.	 A basis vector in this space is:
\begin{eqnarray}
  \left(\begin{array}{cc}
N^{(l)}  \\  (N^{(l)} )^*  \label{Spacegnezero}
\end{array}\right).
\end{eqnarray}
In the case $g=0$ the basis element is $N^{(l)}$. The scalar product is defined in a usual way,
\begin{eqnarray}
\langle \chi | \Psi \rangle =  \int dV   \chi^+ \Psi.
\end{eqnarray}
In case $g \ne 0$ we define the scalar product as
\begin{eqnarray}
\langle  \chi | \Psi \rangle =  \int dV
\left(\begin{array}{cc}
\chi^+ & \Psi^+
 \end{array}\right)
\left(\begin{array}{c}
\chi \\ \Psi 
 \end{array}\right).
\end{eqnarray}
Here $dV$ is an element of volume in $S^2$.

The terms in the expression for the index can be transformed in the following way:

\begin{eqnarray*}
{\tt tr}{\rm  e}^{-t \tilde D^2 } =
 \sum\limits_l \langle n \mid {\rm e}^{-t \tilde D^2} K \mid n \rangle = \\
 \int dV_x dV_y \sum\limits_l (N^{(l)})_{i}(x)^+ G_{ij}(x,y,t) K_{jk}(N^{(l)})_{k}(y) =\\ 
 \int dV_x dV_y G_{ij}(x,y) K_{jk} \delta^{ik}\delta(x-y)&= \\
  \int dV_x G_{ij} (x, x) K_{ji}. \label{tr}
  \end{eqnarray*}
Here $ \mid n \rangle$ is a basis vector which coordinate representation is $ (N^{(l)})^{i}(x)$, $l$ numerates the basis vectors and $i$ numerates the spinor components of each vector. Also we assumed here that
\begin{eqnarray}
\sum\limits_l (N^{(l)})^{k}(y)^+ (N^{(l)})^{i}(x) = \delta(x-y) \delta_{ik} 
\end{eqnarray} 
For $g=0$ it can be reached in a straightforward way. Let us consider the case $g \ne 0$. If we use the representation (\ref{NN^*}) and note that $ x =  \{ \theta_x , \varphi_x \} $ the basis spinors are 
\begin{eqnarray}
 \left(\begin{array}{c}
 \Psi_{nmp}(x) \\ \Psi^*_{nmp}(x)
  \end{array}\right) 
=
 \left(\begin{array}{c}
 e_p P_n^m (\cos \theta_x) e^{im\varphi_x} \\ e_p P_n^m (\cos \theta_x) e^{-im\varphi_x}
  \end{array}\right) 
\end{eqnarray}
Here $e_p$ are basis spinors whose p-th component is equal to 1 and the others are equal to 0; $ P_n^m$ are Legendre polynomials (In this consideration our basis does not have to be a solution of the Dirac equation).

In this basis,
\begin{eqnarray}
\sum\limits_{nmp}  (\Psi_{nmp}^+)^{k}(y)  \Psi_{nmp}^i(x)
 = \delta(\theta_x - \theta_y)
 \left(\begin{array}{cc}
 \delta(\varphi_x - \varphi_y)   &  \delta(\varphi_x + \varphi_y)\\
 \delta(\varphi_x + \varphi_y)  &  \delta(\varphi_x - \varphi_y) 
  \end{array}\right)^{ki}
\end{eqnarray} 
and 
\begin{eqnarray}
 {\tt tr} {\rm e}^{-t \tilde D^2} =
  \int dV_x \left( \sum \limits_{i=1}^{2^{D/2+1}} G_{i,i}(x, x) +  \sum \limits_{i=1}^{2^{D/2}}( G_{i, 2^{D/2}+i}(x,-x) + G_{2^{D/2}+i, i}(x,-x))  \right) K.
\end{eqnarray}
Later (see eq. (\ref{Dsq_expr})) we see that $D^2$ is block-diagonal and therefore (see (\ref{GF}) ) $G$ is block-diagonal.
Consequently, we still can use the expression (\ref{tr}).
Finally,
\begin{eqnarray}
{\tt index} = \int dV_x {\tt tr}(G(x,x,t)K)  \label{IndexExpr}
\end{eqnarray}

Now let us find the index in the case $g=0$.
The expression for $  D_{kin}^2 $ is:
\begin{eqnarray}
  D_{kin}^2  =  \phantom{\{ \partial_{\theta}^2 + \frac{\partial_{\varphi}^2}{\sin^2 \theta} + 
\cot \theta \partial_{ \theta} -\frac{2ie(k+ \kappa\tilde \Gamma^{D+1})A_{\varphi}}{\sin^2 \theta} \partial_{\varphi} +
\Gamma^1 \Gamma^2 \frac{\cos \theta}{\sin^2 \theta}\partial_{\varphi} + C_0  \}}\nonumber\\
- \frac{1}{R^2}   
\left( \partial_{\theta}^2 + \frac{\partial_{\varphi}^2}{\sin^2 \theta} + 
\cot \theta \partial_{ \theta} -\frac{2ie(k+ \kappa\tilde \Gamma^{D}_{(D)})A_{\varphi}}{\sin^2 \theta} \partial_{\varphi} +
\Gamma^1_{(D)} \Gamma^2_{(D)} \frac{\cos \theta}{\sin^2 \theta}\partial_{\varphi} + C_0  \right),
\end{eqnarray}
where
\begin{eqnarray}
C_0 = -\frac{1}{2 \sin^2 \theta} + \frac{\cot^2 \theta}{4} - 
\frac{e^2 A_{\varphi}^2}{\sin^2 \theta} \left(k^2 - \kappa^2 + 2k \kappa \tilde \Gamma^{D}_{(D)}\right) +\nonumber\\
\Gamma^1_{(D)} \Gamma^2_{(D)} \left( \frac{-ie \cos \theta}{\sin^2 \theta}(k + \kappa \tilde \Gamma^{D}_{(D)}) A_{\varphi}  +
 \frac{ie}{\sin \theta} (k + \kappa \tilde \Gamma^{D}_{(D)}) \partial_{\theta} A_{\varphi} \right). \label{D^2}
\end{eqnarray}
In derivation of this expression, we took into account that $A_{\varphi}$ depends only on $\theta$ and does not depend on $\varphi$.
From formula (\ref{GF}) we can find the expression for the Green function of the operator $(-D^2)$. 
In case $x=y$ it is equal to
\begin{eqnarray}
G(x,x,t)= 
 \frac{1}{4\pi R^2} \left( \frac{1}{t}   + \frac{1}{\sin^2 \theta}  (e^2 (k^2 + \kappa^2 +2k \kappa  \tilde \Gamma^{D}_{(D)})A_{\varphi}  + \right.      \nonumber\\ \left.
\frac{ ie}{\sin^2 \theta} \Gamma^1_{(D)} \Gamma^2_{(D)} (k + \kappa  \tilde \Gamma^{D}_{(D)}) A_{\varphi} \cos \theta  )  + C_0  \right)     = \nonumber\\
 \frac{1}{4\pi R^2} \left(     \frac{\cot^2 \theta}{4} -\frac{1}{2 \sin^2 \theta} +   \frac{ie\Gamma^1_{(D)} \Gamma^2_{(D)}}{\sin \theta} (k + \kappa \tilde \Gamma^{D}_{(D)}) \partial_{\theta} A_{\varphi}    \right).
\label{IndEl}
\end{eqnarray}

Since we know $G$, we can now easily find the index by making use of Eq. (\ref{IndexExpr}). The index assosiated with the chirality $K^0$ is equal to
\begin{eqnarray}
{\tt index} = \int d\theta d \varphi R^2 \sin \theta {\tt tr}(G(x,x,t)K) =
 2^{D/2}  \int_{S^2} \frac{ d\theta d \varphi }{4 \pi} ek \partial_{\theta} A_{\varphi}. \label{IndexFormula}
\end{eqnarray}
Since  $A_{\varphi}$ is regular in all points except the South Pole, we can rewrite this expression as
\begin{eqnarray}
{\tt index} =  \frac{ 2^{D/2} e k}{4\pi} \int_L dx^a A_a. \label{IndexFormCurve}
\end{eqnarray}
The last integral is taken over a small curve $L$ surrounding the South Pole . We can calculate it using the properties  (\ref{Eqn/Pg1/1:kurs},\ref{VortexAsymp}). So finally the index assosiated with $K^0$ is equal to
\begin{eqnarray} 
2^{D/2-1}k.\label{FinalIndexG0}
\end{eqnarray}
 The index assosiated with $\tilde K^0$ is calculated in the same way and it is equal to 
\begin{eqnarray} 
2^{D/2-1}\kappa.\label{FinalIndexG0Kappa}
\end{eqnarray}

Let us note that by making use of the equations for the vortex, we can transfrom Eq. (\ref{IndexFormula}) to express the index through the scalar field. The Lagrangian (\ref{VortexLagr}) implies the following field equations: 
 \begin{eqnarray}
\partial_a (\sqrt{G_S} F_{cd} (G_S)^{ac} (G_S)^{bd}) = (G_S)^{ab}( ie \Phi^+ \partial_a \Phi - ie \Phi \partial_a \Phi^+ + 2e^2 A_a | \Phi |^2 ),
\end{eqnarray}
or 
\begin{eqnarray}
A_e = \frac{(G_S)_{be}\partial_a (\sqrt{G_S} F_{cd} (G_S)^{ac} (G_S)^{bd})  }{ 2e^2 |\Phi |^2 } - \frac{i}{2e| \Phi |^2}( \Phi^+ \partial_e \Phi - \Phi \partial_e \Phi^+  ). \label{EOM_vertex}
\end{eqnarray}
Substituting this expression for $A$ into Eq. (\ref{IndexFormCurve}), we obtain
 
\begin{eqnarray}
{\tt index} = \frac{-2^{D/2}ik}{8\pi} \int dx^a \frac{\Phi^+ \partial_a \Phi - \Phi \partial_a \Phi^+}{|\Phi|^2}
\end{eqnarray}
The index assosiated with $\tilde K^0$ is obtained from this expression by the change of $k$ to $\kappa$.

Now let us consider the theory with $g \ne 0$. Like  the previous case,  we find $D^2$:
\begin{eqnarray}
D^2 
  = \phantom{ D_{kin}^2 - g^2 | \Phi|^{4k} \Gamma^0 C (\Gamma^0)^* C^*   and      D_{kin}^2 - g^2 | \Phi|^{4k} \Gamma^0 C (\Gamma^0)^* C^*} \nonumber\\
  \left(\begin{array}{cc}
 D_{kin}^2 - g^2 | \Phi|^{4k} \Gamma^0_{(D)} C_{(D)} (\Gamma^0_{(D)})^* C_{(D)}^*        &
0    \\
0  &
( D_{kin}^*)^2 - g^2 | \Phi|^{4k} (\Gamma^0_{(D)})^* C_{(D)}^* \Gamma^0_{(D)} C_{(D)}   \label{Dsq_expr}
  \end{array}\right)
\end{eqnarray}
In this expression, non-diagonal elements are equal to zero due to Eq. (\ref{FourGammas}).
 In addition one can notice that $ \Gamma^0 C (\Gamma^0)^* C^*  $ is proportional to a unit matrix.
We take into account that ${\tt tr} K^0 = {\tt tr} \tilde{K^0} = 0$ and conclude that this term does not contribute to the index.
Thereby we have the following expression for the index assosiated with $K$:
\begin{eqnarray}
{\tt index} = \int dV_x {\tt tr}(G(x,x,t) K^0  +  G(x,x,t)^* K^0  ) =   2^{D/2}k.
\end{eqnarray}
The index assosiated with the chirality $\tilde{K}$ is calculated in the same way. In cases (b), (c) it is equal to 
\begin{eqnarray}
\int dV_x {\tt tr}(G(x,x,t) \tilde K^0  +  G(x,x,t)^* \tilde K^0  ) =   2^{D/2}\kappa   \label{IndexAnsNoGKappa}
\end{eqnarray}
and in cases (a), (d) it is equal to
\begin{eqnarray}
\int dV_x {\tt tr}(G(x,x,t) \tilde K^0  -  G(x,x,t)^* \tilde K^0  ) =   0. \label{IndexAnsNoG}
\end{eqnarray}
The result in Eq.(\ref{IndexAnsNoG}) can be understood in the following way. By looking at the definition of $\tilde{K}$  (\ref{tildeK}), we see that if this operator acts to a fundamental solution having a form of (\ref{Spacegnezero}), the obtained vector does not have a form of (\ref{Spacegnezero}). For this reason we conclude that in these cases the zero modes are not chiral.

Let us emphasize again that we obtained the index in the case $g \ne 0$ twice as much as in the case $g=0$ only because we calculated them in different spaces. In the first case the space was formed by $N^{(l)}$ and in the second case the space was formed by columns (\ref{Spacegnezero}). If one calculates the index in the case $g = 0$ using the basis (\ref{Spacegnezero})  the obtained answer is  the same as in case $g \ne 0$. Actually this calculation can be done 
 just by taking $g=0$ in the operator $D$.

\section{Conclusions}  \label{Conclusions}
We have considered the Dirac equation with Majorana-like interaction on the vortex on a sphere. We have found the structure of its solutions which is given by (\ref{FinalSolutions}). Also we have calculated two indices of this Dirac operator, which correspond to two its chiralities. In the case $g \ne 0$ they are given by (\ref{IndexAnsNoGKappa}, \ref{IndexAnsNoG}). In the case $g=0$ they are given by (\ref{FinalIndexG0}, \ref{FinalIndexG0Kappa}). In principle in high-dimensional space one can consider some other chiralities and calculate other indices assosiated with them, but we leave this idea for the further job.

Now we consider an example of the case d with $D=6$. In section ~\ref{ZM} the structure of the zero modes is written. By looking at the table \ref{table1} one can check explicitly the index for the case $g=0$. In the case $g \ne 0$  one can check the index assosiated with $K$ by looking to the structure of the solutions described by  (\ref{psi1psi2} -\ref{NlAns}) and the list of non-zero components. Also one can check that the zero modes are not the eigenvectors of $\tilde{K}$. 

The last fact is important from the phenomenological point of view. One can consider a model where the observable world is $M$ and extra dimensions form $S^2$ (see ref. \cite{Frere:2003yv}, \cite{Frere:2010ah}). In this model $\tilde{K}$ is four-dimensional chirality. So if one assume that the zero modes of $N$ are observable particles, they are non-chiral from the four-dimensional point of view, but chiral from the extra-dimensional point of view. This is the difference from the particels formed by Dirac-like interaction, which are chiral from the four-dimensional point of view (see \cite{Frere:2003yv}). 

\section{Acknowledgements}
The author would like to thank his supervisors S.Troitsky , M.Libanov and E.Nugaev for helpful discussions. This work was supported in part by the Dynasty Foundation. This work was supported in part by the grants of the President of the
Russian Federation NS-5525.2010.2, MK-1632.2011.2.

\appendix
\section{Notations}   \label{Notations}

The manifold considered in this paper is  $S^2 \times M $, and $M$ is the Minkovski space. All coordinates of this manifold are numerated by capital Latin letters: $A$, $B$, $\ldots $ .They run from $0$ to $D-1$.

The coordinates on the sphere are $x^1 = \theta$ and $x^2 = \varphi$. $\theta$ is a polar angle and it is counted from the North Pole; $\varphi$ is an axial angle. These coordinates are numerated by small Latin letters: $a$, $b$, $\ldots $ .
So the metric tensor of the sphere is 
\begin{eqnarray}
(G_S)_{ab} = 
R^2 
\left(\begin{array}{cc}
1 & 0\\
0 & \sin^2 \theta 
\end{array}\right)  
\end{eqnarray}
Its determinant is designated by $G_S$:
$$  G_S  = {\rm det}(G_S)_{ab}. $$

The coordinates of $M$ are numerated by Greek letters: $\mu$, $\nu$, $\ldots $ . The zero coordinate $x^0$ is time. Spatial coordinates are numerated by Latin letters: $i$, $j$, $\ldots $ .The metric tensor for the whole manifold is
\begin{eqnarray}
G_{AB}= {\rm diag} (\begin{array}{ccccccc}
1 & -(G_S)_{ab} & -1 & \ldots
 -1
\end{array}  )
\end{eqnarray}.

In the case $D=3$ gamma-matrices are defined as:
\begin{eqnarray}
\Gamma^0_{(2)} =  \sigma_3, \quad 
\Gamma^1_{(2)} =- i\sigma_2, \quad
\Gamma^2_{(2)} =
 i\sigma_1.\label{2D_gamma_matrices}
\end{eqnarray}
The charge conjugation matrix is $C = \Gamma^2_{(2)}$. Here $\sigma^1$, $\sigma^2$, $\sigma^3$ are Pauli matrices.
In this paper we also consider $\Gamma^0_{(2)} = i \Gamma^1_{(2)} \Gamma^2_{(2)}$ as chirality operator.

Let us construct gamma-matrices in higher-dimensional spaces. If the number of dimensions is even and equal to $D$, we can define gamma-matrices $\Gamma^{A}_{(D)}$  by using $(D-2)$-dimensional gamma-matrices $\gamma^{A}_{(D-2)}$ in the following way:
\begin{eqnarray}
\Gamma^{A}_{(D)} &=& 
\left(\begin{array}{cc}
0 & \Gamma^{A}_{(D-2)} \\
\Gamma^{A}_{(D-2)} & 0 
\end{array}\right),  \quad A = 0, \ldots , D-2; \label{Gamma_matrices_induction}
  \\
\Gamma^{D-1}_{(D)} &=& 
\left(\begin{array}{cc}
0 & -1\\
1 & 0 
\end{array}\right).  \label{HDGammamatrices}
\end{eqnarray}

If we want to consider the case of odd number of dimensions which is equal to $D+1$ we also introduce
\begin{eqnarray}
\Gamma^{D}_{(D)} = i
\left(\begin{array}{cc}
1 & 0\\
0 & -1 
\end{array}\right). 
\end{eqnarray}

Also we introduce a projection operator $ \tilde \Gamma^{D}_{(D)} = {\rm diag}
\left(\begin{array}{ccc}
1  &,& -1 
\end{array}\right) 
 $ . One can check that 
\begin{eqnarray}  
\tilde \Gamma^{D}_{(D)} =-i \Gamma^{D}_{(D)} = 
(-i)^{\frac{D}{2}+1}   \Gamma^0_{(D)} * \ldots *\Gamma^{D-1}_{(D)}.
\end{eqnarray}

One can notice that our gamma-matrices are imaginary if $A = 2,4,\ldots , D$(any even number more than 0) and real otherwise. In the  D-dimensional space we can define charge conjugation matrices as
\begin{eqnarray}
C^{(1)}_{(D)} = \Gamma^{2}_{(D)} \Gamma^{4}_{(D)}...\Gamma^{D}_{(D)}, \\
C^{(2)}_{(D)} = \Gamma^{2}_{(D)} \Gamma^{4}_{(D)}...\Gamma^{D-2}_{(D)}.
\end{eqnarray}
In the $D+1$-dimensional space only $C^{(1)}_{(D)}$ is suitable.

Defined in this way, $C$-matrices have the following properties in $D \ge 4$:
\begin{eqnarray}
\Gamma^A_{(D)} C^{(2)}_{(D)} =(-1)^{ \frac{D}{2} - 1 } C^{(2)}_{(D)} (\Gamma^A_{(D)})^*,  \\
\Gamma^A_{(D)} C^{(1)}_{(D)} =(-1)^{ \frac{D}{2}      } C^{(1)}_{(D)} (\Gamma^A)^*. \label{Gamma_C}
\end{eqnarray}

 Matrix $C_{(D)}$ written without upper index denotes any of these charge conjugation matrices.

By making use of the last formulae, one can show that
 \begin{eqnarray}
\Gamma^0_{(D)} \Gamma^A_{(D)} \Gamma^0_{(D)} C_{(D)} + \Gamma^0_{(D)} C_{(D)} ( \Gamma^0_{(D)})^* (\Gamma^A_{(D)})^* = 0. \label{FourGammas}
\end{eqnarray}
This relation is useful in derivation of many of the expressions mentioned in this paper.

One can show that 
\begin{eqnarray}
(\Gamma^0_{(D)} C^{(1)}_{(D)})^T = (-1)^{\frac{D}{4}(\frac{D}{2}+1)}\Gamma^0_{(D)} C^{(1)}_{(D)},\nonumber\\
(\Gamma^0_{(D)} C^{(2)}_{(D)})^T = (-1)^{\frac{D}{4}(\frac{D}{2}-1)}\Gamma^0_{(D)} C^{(2)}_{(D)},\label{(Gamma^0_{(D)} C_{(D)})^T}\\
(C^{(2)}_{(D)})^{-1} = (-1)^{ \frac{D}{4}( \frac{D}{2} -1 ) } C^{(2)}_{(D)},\nonumber\\
(C^{(1)}_{(D)})^{-1} = (-1)^{ \frac{D}{4}( \frac{D}{2} +1 ) } C^{(1)}_{(D)},\\
(C^{(1)}_{(D)})^* C^{(1)}_{(D)} = (-1)^{\frac{D}{4}(\frac{D}{2}+3)},\label{CC_star}\\
(C^{(2)}_{(D)})^* C^{(2)}_{(D)} = (-1)^{(\frac{D}{4}+1)(\frac{D}{2}-1)}.
\end{eqnarray}

Also in this paper we use the properties:
\begin{eqnarray}
\{ \Gamma^{D}_{(D)},  \Gamma^0_{(D)} C_{(D)}  \}              &=& 0 \quad \mbox{if D/2 mod 4 = 1 or 3}  \nonumber\\
\left[ \Gamma^{D}_{(D)},  \Gamma^0_{(D)} C_{(D)}  \right]  &=& 0 \quad  \mbox{if D/2 mod 4 = 2 } \\
\left[ \Gamma^{D}_{(D)},   C_{(D)}  \right]              &=& 0 \quad \mbox{if D/2 mod 4 = 1 or 3}  \nonumber\\
\{ \Gamma^{D}_{(D)},   C_{(D)}  \}  &=& 0 \quad  \mbox{if D/2 mod 4 = 2 }  \label{CommutRel}
\end{eqnarray}

 The structure of $C$-matrices in the cases used in this paper is the following: 
\begin{itemize}
\item{(a)} 
D/2 mod 4 = 1,
\begin{eqnarray}
C^{(1)}_{(D)} = i \left(\begin{array}{cc}
C^{(1)}_{(D-2)} & 0\\
0 & -C^{(1)}_{(D-2)} 
\end{array}\right),   
\end{eqnarray} 

 \item{(b)}
D/2 mod 4 = 2,
\begin{eqnarray}
 C^{2}_{(D)} = \left(\begin{array}{cc}
0 & C^{(1)}_{(D-2)}\\
C^{(1)}_{(D-2)} & 0 
\end{array}\right),  
\end{eqnarray}

 \item{(c)}
D/2 mod 4 = 2,
\begin{eqnarray}
C^{(1)}_{(D)} = i \left(\begin{array}{cc}
0 & -C^{(1)}_{(D-2)}\\
C^{(1)}_{(D-2)} & 0 
\end{array}\right),  
\end{eqnarray}

\item{(d)}  
D/2 mod 4 = 3, 
\begin{eqnarray}
C^{(2)}_{(D)} =  \left(\begin{array}{cc}
C^{(1)}_{(D-2)} & 0\\
0 & C^{(1)}_{(D-2)} 
\end{array}\right).   
\end{eqnarray} 

\end{itemize}


\section{The solution of the equations for zero modes} \label{solving}

Here we explore the fundamental solutions of the equations (\ref{2D_Dirac_eq_1} - \ref{2D_Dirac_eq_2}). In the case $g=0$ we find them explicitly and in the case $g \ne 0$ we find their asymptotics. We assume $k > 0$. 
 Let us write these equations in the component form:

\begin{eqnarray}
(\partial_{\theta} + \frac{\cot \theta}{2}  ) \xi^2_{n}  -         \frac{i}{\sin \theta}(\partial_{\varphi} - ie(\kappa+k)A_{\varphi} )\xi^2_{n} -         igR \Phi^{2k} \eta^1_{N+1-n}=0,   \nonumber \\
(\partial_{\theta} + \frac{\cot \theta}{2}  ) \xi^1_{n}  +         \frac{i}{\sin \theta}(\partial_{\varphi} - ie(\kappa+k)A_{\varphi} )\xi^1_{n} +        igR \Phi^{2k} \eta^2_{N+1-n}=0,   \nonumber\\
(\partial_{\theta} + \frac{\cot \theta}{2}  ) (\eta^1_{N+1-n}) + \frac{i}{\sin \theta}(\partial_{\varphi} - ie(\kappa-k)A_{\varphi} )(\eta^1_{N+1-n}) + igR (\Phi^*)^{2k} \xi^2_n =0,   \nonumber\\
(\partial_{\theta} + \frac{\cot \theta}{2}  ) (\eta^2_{N+1-n}) - \frac{i}{\sin \theta}(\partial_{\varphi} - ie(\kappa-k)A_{\varphi} )(\eta^2_{N+1-n}) - igR (\Phi^*)^{2k} \xi^1_n =0;   \label{CompEq4} 
\end{eqnarray}

To get rid of the dependence on $\varphi$ we use the following ansatz: 
\begin{eqnarray}
\xi^2_n = u_L \exp{\left(-i(l+\frac{1}{2})\varphi + \int_0^{\theta} d \theta \frac{eA_{\varphi}}{\sin \theta}( \kappa + k)\right)}, \nonumber\\
\xi^1_n = u_U \exp{\left( i(l+\frac{1}{2})\varphi + \int_0^{\theta} d \theta \frac{eA_{\varphi}}{\sin \theta}(-\kappa - k)\right)}, \nonumber\\
\eta^1_{N+1-n} = v_L \exp{\left( -i(2k+l+\frac{1}{2})\varphi + \int_0^{\theta} d \theta \frac{eA_{\varphi}}{\sin \theta}(-\kappa + k)\right)},\nonumber\\
\eta^2_{N+1-n} = v_U \exp{\left( -i(2k-l-\frac{1}{2})\varphi + \int_0^{\theta} d \theta \frac{eA_{\varphi}}{\sin \theta}(\kappa - k)\right)}. \label{uv} 
\end{eqnarray}
Here $u_L, u_U, v_L, v_U$ are new variables depending only on $\theta$, and $l$ is an integer number which further will numerate the solutions. Let us note that in this ansatz, we assumed that the spinor components obey antiperiodical boundary conditions. They are chosen because if we move on a closed curve from $\varphi = 0$ to $\varphi = 2\pi$ we rotate our velbain (\ref{velbain}) by $2\pi$. This results in the change of the sign of a fermion.
In terms of the new variables the equations read
\begin{eqnarray}
\partial_{\theta}u_L - \frac{l+1/2}{\sin \theta} u_L + \frac{\cot}{2} u_L - gR f^{2k}e^{-2\int \frac{e \kappa A_{\varphi} d \theta}{\sin \theta}} i  v_L = 0, \nonumber \\
\partial_{\theta}u_U - \frac{l+1/2}{\sin \theta} u_U + \frac{\cot}{2} u_U + gR f^{2k}e^{ 2\int \frac{e \kappa A_{\varphi} d \theta}{\sin \theta}} i v_U = 0, \nonumber \\
\partial_{\theta}v_L + \frac{2k+l+1/2}{\sin \theta} v_L + \frac{\cot}{2} v_L + gR f^{2k}e^{ 2\int \frac{e \kappa A_{\varphi} d \theta}{\sin \theta}} i  u_L = 0, \nonumber\\
\partial_{\theta}v_U - \frac{2k-l-1/2}{\sin \theta} v_U + \frac{\cot}{2} v_U - gR f^{2k}e^{-2\int \frac{e \kappa A_{\varphi} d \theta}{\sin \theta}} i u_U = 0. 
\end{eqnarray}

Then we make another change:
\begin{eqnarray}
u_L = y_L \frac{(\sin \theta/2)^l }{(\cos \theta/2)^{l+1}} , \nonumber\\
v_L = z_L \frac{(\sin \theta/2)^{-2k-l-1} }{(\cos \theta/2)^{-2k-l}}, \nonumber\\
u_U = y_u \frac{(\sin \theta/2)^l }{(\cos \theta/2)^{l+1}}, \nonumber\\
v_U = z_u \frac{(\sin \theta/2)^{2k-l-1} }{(\cos \theta/2)^{2k-l}} \label{yz}
\end{eqnarray}

to we obtain the following equations:
\begin{eqnarray}
\partial_{\theta}y_L - igR f^{2k}e^{-2\int \frac{e \kappa A_{\varphi} d \theta}{\sin \theta}} (\tan \theta/2)^{-2k-2l-1} z_L = 0, 
 \\
\partial_{\theta}z_L + igR f^{2k}e^{ 2\int \frac{e \kappa A_{\varphi} d \theta}{\sin \theta}} (\tan \theta/2)^{2k+2l+1} y_L = 0, \nonumber\\
\partial_{\theta}y_U + igR f^{2k}e^{ 2\int \frac{e \kappa A_{\varphi} d \theta}{\sin \theta}} (\tan \theta/2)^{2k-2l-1} z_U = 0, \nonumber\\
\partial_{\theta}z_U - igR f^{2k}e^{-2\int \frac{e \kappa A_{\varphi} d \theta}{\sin \theta}} (\tan \theta/2)^{-2k+2l+1} y_U = 0. \label{eq_z_U}
\end{eqnarray}

Now let us consider separately the case $g=0$. Here we conclude that all variables in the equations (\ref{eq_z_U}) are constants:
\begin{eqnarray}
y_L = y_L^0 = {\rm const}, \nonumber\\ 
y_u = y_u^0 = {\rm  const}, \nonumber \\
z_L = z_L^0 =  {\rm  const} , \nonumber \\
z_U = z_U^0 = {\rm  const} .
\end{eqnarray}
Substituting them into the expressions  (\ref{uv} , \ref{yz}) we find the solutions for the initial variables:
\begin{eqnarray}
\xi^2_n = y_L^0 \frac{(\sin \theta/2)^l }{(\cos \theta/2)^{l+1}} \exp{\left(-i(l+\frac{1}{2})\varphi + \int_0^{\theta} d \theta \frac{eA_{\varphi}}{\sin \theta}( \kappa + k)\right)} , \nonumber\\
\xi^1_n = y_U^0 \frac{(\sin \theta/2)^l }{(\cos \theta/2)^{l+1}} \exp{\left( i(l+\frac{1}{2})\varphi + \int_0^{\theta} d \theta \frac{eA_{\varphi}}{\sin \theta}(-\kappa - k)\right)}, \nonumber\\
\eta^1_{N+1-n} = z_L^0  \frac{(\sin \theta/2)^{-2k-l-1} }{(\cos \theta/2)^{-2k-l}}  \exp{\left( -i(2k+l+\frac{1}{2})\varphi + \int_0^{\theta} d \theta \frac{eA_{\varphi}}{\sin \theta}(-\kappa + k)\right)}  , \nonumber\\
\eta^2_{N+1-n} =  z_u^0  \frac{(\sin \theta/2)^{2k-l-1} }{(\cos \theta/2)^{2k-l}}  \exp{\left( -i(2k-l-\frac{1}{2})\varphi + \int_0^{\theta} d \theta \frac{eA_{\varphi}}{\sin \theta}(\kappa - k)\right)}. \label{RawAns0}
\end{eqnarray}
These expressions may have singularities at $\theta \to 0$ or $\theta \to \pi$. To investigate them we write the asymptotics of Eqs. (\ref{RawAns0}) which are the following ($\xi = \pi - \theta$):
\begin{itemize}
\item{ $\theta \to 0$:}
\begin{eqnarray} 
\xi^2_n \sim y_L^0 (\theta/2)^l \exp{\left(-i(l+\frac{1}{2})\varphi\right)}, \nonumber\\
\xi^1_n \sim y_U^0 (\theta/2)^l \exp{\left( i(l+\frac{1}{2})\varphi\right)}, \nonumber\\
\eta^1_{N+1-n} \sim z_L^0 (\theta/2)^{-2k-l-1}\exp{\left( -i(2k+l+\frac{1}{2})\varphi\right)},\nonumber\\
\eta^2_{N+1-n} \sim z_u^0  (\theta/2)^{2k-l-1}  \exp{\left( -i(2k-l-\frac{1}{2})\varphi\right)}.  \nonumber
\end{eqnarray}

\item{$\theta \to \pi$:}
\begin{eqnarray}
\xi^2_n \sim y_L^0 \zeta^{-k-\kappa-l-1} \exp{\left(-i(l+\frac{1}{2})\varphi\right)}, \nonumber\\
\xi^1_n \sim y_U^0 \zeta^{\kappa+k-l-1} \exp{\left( i(l+\frac{1}{2})\varphi\right)}, \nonumber\\
\eta^1_{N+1-n} \sim z_L^0 \zeta^{\kappa+k+l}\exp{\left( -i(2k+l+\frac{1}{2})\varphi\right)},\nonumber\\
\eta^2_{N+1-n} \sim z_u^0  \zeta^{-\kappa-k+l}  \exp{\left( -i(2k-l-\frac{1}{2})\varphi\right)}.  \label{NtoPi}
\end{eqnarray}
\end{itemize}
 $\zeta = \pi - \theta$

These expressions are regular if $\theta$ and $\zeta$ have non-negative powers.
From this condition we find the constraints for $l$:
\begin{eqnarray}
\xi^2_n : 0 \le l \le -k-\kappa-1, \nonumber\\
\xi^1_n : 0 \le l \le  k+\kappa-1, \nonumber \\
\eta^1_{N+1-n} : -\kappa-k \le l \le -2k-1, \nonumber\\
\eta^2_{N+1-n} : \kappa +k \le l \le 2k-1. \nonumber
\end{eqnarray}

Finally we can write the structure of the solutions depending on $k$ and $\kappa$ (recall that $k>0$):

\begin{description}
\item[$\kappa+k \le -1$]
$\xi^2_n$ has $|\kappa+k|$ modes and $\eta^2_{N+1-n}$ has $k-\kappa$ modes. 
\item[$\kappa+k = 0 $]
Only $\eta^2_{N+1-n}$ has non-zero modes. 
Their number is $2k$.
\item[$1 \le \kappa+k \le 2k-1$]
$\xi^1_n$ has $\kappa+k$ modes and $\eta^2_{N+1-n}$ has $k - \kappa$ modes. 
\item[$\kappa+k = 2k$]
$\xi^1_n$ has $2k$ non-zero modes.
\item[$\kappa+k \ge 2k+1$]
$\xi^1_n$ has $\kappa+k$ modes and $\eta^1_{N+1-n}$ has $\kappa - k$ modes. 
\end{description}

Now let us consider the case $g \ne 0$. A general analytical solution to  the equations (\ref{eq_z_U}) is unknown. In this paper we find the asymptotics of the solutions in cases $\theta \to 0$ and $\theta \to \pi$. To do it we derive the second-order equations containing only $y_L$, $y_U$ from Eqs. (\ref{eq_z_U}):
\begin{eqnarray}
  \partial_{\theta}^{2}y_{L}   +   \left(\frac{1+2k+2l}{\sin\theta} - 2k\frac{\partial_{\theta}f}{f} +\frac{2e\kappa A}{\sin \theta}\right) \partial_{\theta}y_{L}  -  (gR)^{2}f^{4k}y_{L}  =  0, \nonumber\\
  \partial_{\theta}^{2}y_{U}   +   \left(\frac{1-2k+2l}{\sin\theta} - 2k\frac{\partial_{\theta}f}{f} - \frac{2e\kappa A}{\sin \theta}\right) \partial_{\theta}y_{U}  -  (gR)^{2}f^{4k}y_{U}  =  0. \label{SOE}
    \end{eqnarray} 
 Each of these equations has two linearly independent solutions. After substituting the asymptotics of the vortex fields, eq. (\ref{VortexAsymp}), we can find the behavior of the unknown functions:  

\begin{itemize} 
\item{ $\theta \to 0$:} 
\begin{eqnarray}  
  y_{L} \sim 2^{l}c_{0} \left(1 + \frac{g^2R^2C_{0F}^{4k}}{2(4k+2l+2)(2k+1)} \theta^{4k+2}\right)  +   2^{l}d_{0}g \theta^{-2l}, \nonumber\\
   y_{U} \sim 2^{l}a_{0} \left(1 +  \frac{g^2R^2C_{0F}^{4k}}{2(2l+2)(2k+1)}\theta^{4k+2}\right)     +    2^{l} b_{0} g \theta^{4k-2l}. \nonumber
\end{eqnarray} 
\item{ $\theta \to \pi$:}  
\begin{eqnarray}
y_{L} \sim 2^{-l-1} c_{\pi} \left(1 - \frac{(gR)^{2}C_{\pi F}^{4k}\zeta^{2}}{4(k+\kappa +l)}\right) + 2^{-l-1} d_{\pi} g \zeta^{2l+2k+2\kappa +2}, \nonumber\\
y_{U} \sim 2^{-l-1} a_{\pi} \left(1 + \frac{(gR)^{2}C_{\pi F}^{4k}\zeta^{2}}{4(k+\kappa -l)}\right) + 2^{-l-1} b_{\pi} g \zeta^{2l-2k-2\kappa +2}.
\end{eqnarray}  
\end{itemize}
Here $a_0,  b_0 , c_0 , d_0, a_{\pi} , b_{\pi} , c_{\pi} , d_{\pi}$ are arbitrary constants and factors $2^l$, $2^{-l-1}$ and $g$ are introduced here for convenience.

If we substitute the last expressions into Eqs. (\ref{eq_z_U}, \ref{uv} , \ref{yz}), we find the behavior of our initial functions:

\begin{itemize}
 \item{ $\theta \to 0$:}  
\begin{eqnarray}
\xi^2_n \sim \left( c_{0}\theta^{l} + d_{0}g\theta^{-l} \right){\rm e}^{-i(l+1/2)\varphi}, \nonumber\\
\xi^1_n \sim  \left(a_{0}\theta^{l} + b_{0}g\theta^{4k-l}\right){\rm e}^{i(l+1/2)\varphi}, \nonumber \\
\eta^1_{N+1-n} \sim  \left(\frac{-ic_{0}gRC_{0F}^{2k}\theta^{2k+l+1}}{4k+2l+2}      + \frac{id_{0}(2l)}{RC_{0F}^{2k}}\theta^{-2k-l-1}\right){\rm e}^{-i(2k+l+1/2)\varphi}, \nonumber\\
\eta^2_{N+1-n} \sim \left(\frac{igRC_{0F}^{2k}a_{0}\theta^{2k+l+1}}{2l+2}        + \frac{ib_{0}(4k-2l)}{RC_{0F}^{2k}}\theta^{2k-l-1}\right){\rm e}^{-i(2k-l-1/2)\varphi}. \nonumber
\end{eqnarray}
\item{ $\theta \to \pi$: }
\begin{eqnarray}
\xi^2_n \sim \left( c_{\pi}\zeta^{-k-\kappa-l-1}    +   d_{\pi}g\zeta^{l+k+\kappa+1} \right) {\rm e}^{-i(l+1/2)\varphi}, \nonumber\\ 
\xi^1_n \sim  \left( a_{\pi} \zeta^{k+\kappa-l-1}    +b_{\pi} g\zeta^{l-k-\kappa+1}\right){\rm e}^{i(l+1/2)\varphi},\nonumber \\
\eta^1_{N+1-n} \sim \left( \frac{-i gRC_{\pi F}^{2k}c_{\pi}\zeta^{-k-\kappa-l}}{2l+2k+2\kappa}      +\frac{i d_{\pi}(2l+2k+2\kappa +2)\zeta^{l+k+\kappa}}{RC_{\pi F}^{2k}  }\right) {\rm e}^{-i(2k+l+1/2)\varphi}, \nonumber\\
\eta^2_{N+1-n} \sim  \left(\frac{-i gRC_{\pi F}^{2k}a_{\pi}}{-2l+2k+2\kappa}\zeta^{k+\kappa-l}+ \frac{-i b_{\pi}}{RC_{\pi F}^{2k}}(2l-2k-2\kappa+2)\zeta^{l-k-\kappa}\right){\rm e}^{-i(2k-l-1/2)\varphi}. \label{Ans_N_4}
\end{eqnarray}
\end{itemize}

To select regular solutions we use the so-called different signs theorem (see also \cite{Frere:2003yv}, \cite{Libanov:2000uf}). It states that if $y_{L,U}$ and  $\partial_{\theta}y_{L,U}$ have the same sign at one interior point $\theta_0$ of the interval $(0 , \pi)$, then they have the same sign at all points of the interval $ (\theta_0 , \pi)$. Let us prove it. If a function and its derivative have the same sign which is positive(negative) the function cannot change it before its derivative does it. The derivative can change its sign in a point of a local maximum(minimum). From Eqs.(\ref{SOE}) we see that if  $\partial_{\theta}y_{L,U} = 0$ and $\partial_{\theta}f/f$, $A_{\varphi}$ do not have any singularities (this is true for the vortex fields), then  $y_{U,L}$ and $\partial^2_{\theta}y_{U,L}$ have the same sign. Thus we got the contradictory statement that in the point of a local maximum(minimun) the second derivative is positive(negative).

Now let us consider the solutions $\xi^2_n$ and $\eta^1_{N+1-n}$. One can see that their linearly independent components have opposite powers of $\zeta$ in case $\theta \to \pi$ so only one of them should be non-zero if we want to get a regular expression. So there are two possible cases:

\begin{eqnarray}
-k-\kappa-l-1 \ge 0 :\quad
 c_{\pi} \ne 0 , d_{\pi}=0, \nonumber\\
 k+\kappa+l \ge 0 : \quad
 c_{\pi} = 0 , d_{\pi}\ne 0.
\end{eqnarray}
Since the asymptotics have the power-like form, we conclude that in both cases if $\zeta$ increases, $y_L(\zeta)$ also increases.
If we want $\xi^2_n$ to be regular, only one of the coefficients $c_0$,$d_0$ should be non-zero. If we suppose that $d_0 \ne 0$, then from the expression for $\eta^1_{N+1-n}$ we conclude that $l < -2k-1$ and therefore $y_L(\theta)$ increases. This contradicts to the different signs theorem. If we suppose that $c_0 \ne 0$, from the expression for $\xi^2_n$ we conclude that $l \ge 0$ and so $y_L(\theta)$ also increases if $\theta$ increases. This possibility also contradicts to the different signs theorem.

The case of $\xi^1_n$ and $\eta^2_{N+1-n}$ is treated in the same way.
From the asymptotics of $\theta \to \pi$ we see that there are two cases:
\begin{eqnarray}
k+\kappa-l-1 \ge 0 : \quad
a_{\pi} \ne 0 , b_{\pi}=0, \nonumber\\
l-k-\kappa \ge 0 :\quad
a_{\pi} = 0 , b_{\pi} \ne 0.
\end{eqnarray}

In these cases (as opposed to the previous ones) we concede that  $a_0,b_0 \ne 0$ together. From the condition that $\xi^1_n, \eta^2_{N+1-n}$ cannot have negative powers of $\theta, \xi$, we conclude that $0 \le l \le 2k-1$. Also, if we force these coefficients to have different signs, we can satisfy the different signs theorem.
Let us consider the case when only one of the coefficients  $a_0,b_0 $  is non-zero. For $a_0 \ne 0$ we have $l > 0 $ and so $y_U(\theta)$ does not satisfy the theorem; and for $b_0 \ne 0$ we have $ l < 2k-1$ and $y_U(\theta)$ again does not satisfy the different signs theorem.

Thus we have shown that  $\xi^1_n$ and $ \eta^2_{N+1-n}$ have regular solutions but  $\xi^2_n$ and $\eta^1_{N+1-n}$ do not have them.

One can note that the solutions in the cases $g = 0$ and $g \ne 0$ have different structure. Nevertheless, if we take the expressions (\ref{Ans_N_4}) and find their limit in case $g \to 0$, we obtain the expressions having the same form as Eqs.(\ref{NtoPi}).


\section{The Green function of the Heat equation} \label{A_I}
Let us consider a general Heat equation
\begin{eqnarray}
 \frac{d}{dt} G(x,y,t)     = ( a^{\alpha \beta}\partial_{\alpha}\partial_{\beta} + b^{\alpha}\partial_{\alpha} + c) G(x,y,t) \label{GenOp}
\end{eqnarray} 
supplemented by the condition
\begin{eqnarray}
  G(x,y,t) \to \delta(x-y)     \qquad \mbox{at} \quad  t \to 0. \label{BC}
\end{eqnarray}
Here $\partial_{\alpha} = \frac{\partial}{\partial x^{\alpha}}$ and $a$, $b$, $c$ depend on $x^{\alpha}$.
We can solve this equation in the case $x \approx y$ and $t \to 0$   
  with the following ansatz:
\begin{eqnarray}
 G(x,y,t) = e^{S(x,y)/t} \sum\limits_{k=-1} A_{k}(x,y)t^{k}.  \label{Gansatz}
 \end{eqnarray}

By substituting it into Eq.(\ref{GenOp}) and equating the terms having equal powers of $t$, we get the following equations:
\begin{eqnarray} 
&& -S = a^{\alpha \beta}\partial_{\alpha}S \partial_{\beta}S,
\label{k=-3}
\\ 
&& A_{-1} = -a^{\alpha \beta} \partial_{\alpha}\partial_{\beta}S A_{-1} -2a^{\alpha \beta}\partial_{\alpha}S\partial_{\beta}A_{-1}-b^{\alpha}\partial_{\alpha}SA_{-1},
\label{k=-2}
\\ 
&& 0 = a^{\alpha \beta}\partial_{\alpha}\partial_{\beta}S A_{0} + 2a^{\alpha \beta}\partial_{\alpha}S \partial_{\beta} A_{0}  +  b^{\alpha}\partial_{\alpha}SA_{0} +
 \nonumber\\ && \phantom{ 0 = a^{\alpha \beta}\partial_{\alpha}\partial_{\beta}S A_{0} + 2a^{\alpha \beta}\partial_{\alpha}S \partial_{\beta} A_{0} }
(a^{\alpha \beta}\partial_{\alpha}\partial_{\beta} + b^{\alpha}\partial_{\alpha}+c)A_{-1}.
\label{k=-1}
\end{eqnarray}

Now we take into account that  $x \approx y$ and assume $a,b,c(x) \approx a,b,c(y)$.
The equation (\ref{k=-3}) has the solution
\begin{eqnarray}
S = -\frac{1}{4} a_{\alpha \beta} \Delta x^{\alpha} \Delta x^{\beta}, \label{Sxy}
\end{eqnarray}
where 
$$ a_{\alpha \beta} = (a^{\alpha \beta})^{-1}, $$
$$ \Delta x^{\alpha} \equiv x^{\alpha} - y^{\alpha}. $$
Using the expression for $S$ we transform the Eq. (\ref{k=-2}) to
\begin{eqnarray}
\Delta x^{\alpha} \partial_{\alpha} A_{-1} + \frac{b^{\alpha} a_{\gamma \alpha} \Delta x^{\gamma}}{2} A_{-1} = 0.
\end{eqnarray}
It's solution is 
\begin{eqnarray}
A_{-1} = e^{-\frac{b^{\alpha} a_{\gamma \alpha} \Delta x^{\gamma}}{2}} \label{A-1}
\end{eqnarray}
By substituting $S$ and $A_{-1}$ into Eq. (\ref{k=-1}), we get
\begin{eqnarray}
0 = -A_0 - \Delta x^{\beta} \partial_{\beta} A_0  - \frac{b^{\alpha} a_{\gamma \alpha} \Delta x^{\gamma}}{2} A_0 +(-\frac{b^{\alpha} b^{\beta} a_{\alpha \beta}}{4}+c)A_{-1},
\end{eqnarray}
Its general solution is 
\begin{eqnarray}
A_0 = \left(-\frac{b^{\alpha} b^{\beta} a_{\alpha \beta}}{4}+c + \frac{ {\rm const} }{\sqrt{ \sum\limits_{\alpha} (\Delta x_{\alpha})^2 }}\right)A_{-1}  \label{A-0}
\end{eqnarray}
To avoid a singularity in the last expression we take ${\rm const} = 0$.

Finally after substituting (\ref{Sxy}), (\ref{A-1}), (\ref{A-0}) into (\ref{Gansatz}) we have
\begin{eqnarray}
G = \frac{1}{4\pi } \left(\frac{1}{t} - \frac{b^{\alpha} a_{\alpha \beta}b^{\beta}}{4}+c  \right)  e^{-\frac{1}{4t} a_{\alpha \beta} \Delta x^{\alpha} \Delta x^{\beta}  -\frac{b^{\alpha} a_{\gamma \alpha} \Delta x^k}{2}}  \label{GF}
\end{eqnarray}
The constant $\frac{1}{4\pi }$ is introduced here to satisfy the condition (\ref{BC}).


\end{document}